\documentstyle[preprint,aps,epsf]{revtex}
\def\bbox#1{\mbox{\boldmath$#1$}}

\begin{document}
\vspace*{-2cm}
\begin{center}
{\Large \bf Exact two-loop vacuum polarization correction
\vspace{0.1cm}

to the Lamb shift in hydrogen-like ions}
\end{center}
\normalsize
\bigskip
\begin{center}

\vspace{0.3cm}
{\large G\"unter Plunien, Thomas Beier and Gerhard Soff}

{\sl Institut f\"ur Theoretische Physik, Technische Universit\"at Dresden,\\
Mommsenstr. 13, D-01062 Dresden, Federal Republic of Germany} 

\vspace{0.3cm}
{\large Hans Persson}

\vspace{0.2cm} 
{\sl Department of Physics, Chalmers University of Technology and \\
the University of Gothenburg, S-412 96 G\"oteborg, Sweden}

\end{center}

\vspace{1.0cm}
\begin{abstract}
We present a calculation scheme for the two-loop vacuum polarization
correction of order $\alpha^2$ to the Lamb shift of hydrogen-like 
high-$Z$ atoms. The interaction with the external Coulomb field is taken 
into account to all orders in $(Z\alpha)$. By means of a modified potential 
approach the problem is reduced to the evaluation of effective one-loop 
vacuum polarization potentials. An expression for the energy shift is deduced 
within the framework of partial wave decomposition performing appropriate 
subtractions. 
Exact results for the two-loop vacuum polarization contribution to the
Lamb shift of K- and L-shell electron states in hydrogen-like Lead and 
Uranium are presented.
\end{abstract}

\bigskip
PACS-numbers: 31.10.+z, 31.30.-i, 31.30.Jv


\newpage

\section{Introduction}
Recent experimental progress in the spectroscopy of highly charged heavy 
ions \cite{sea91,bey94,sea93,bey95} demands theoretical predictions for the
Lamb shift, which should include the complete set of QED radiative
corrections of order $\alpha^2$ but accounting for all orders $(Z\alpha)$ in
the interaction with the strong external Coulomb field. For low-$Z$ 
elements, a potential expansion with respect to powers in $(Z\alpha)$
is legitimate and all $\alpha^2$-corrections have recently been calculated 
up to the order of $\alpha^2 (Z\alpha)^5$ \cite{eides,pachucki}. 
However, for systems under consideration in recent Lamb-shift measurements
such as Gold or Uranium, a value for the effective coupling 
$Z\alpha > 0.5$ already indicates that $Z\alpha$-expansion in the regime
of large $Z$ becomes inadequate. 
Instead, exact electron propagators and wave functions in the external
Coulomb field of extended nuclei have to be used in calculations of 
all second-order diagrams (Fig. \ref{f:1}). 
Meanwhile, calculational approaches are available for most of these QED 
effects (see the work of Persson {\em et al.} \cite{pea96} and cited 
references). 
However, the calculations are not yet complete: 
1.~Exact numerical evaluations of the complete set of two-photon 
self-energy contributions remains as a major challenge, although recent
progress has been made in deriving renormalized expressions for the 
resulting energy shifts \cite{lam95}. 
The lack of numerical results for these contributions
represents a major uncertainy in theoretical predictions for the 
Lamb shift aiming for a relative precision of $10^{-6}$ for the total
electron binding energy.
2.~Exact evaluation schemes, which also treat the loops involved to all 
orders in $(Z\alpha)$, have been developed \cite{pwr} only for the combined 
self energy - vacuum polarization corrections SEVPabc and the two-loop ladder 
vacuum polarization diagram VPVPa (see Fig. \ref{f:1}). 
Values for the VPVPa correction for large $Z$-numbers have been tabulated 
recently \cite{bea97}. The self energy-vacuum polarization S(VP)E 
(Fig. \ref{f:1}) is calculated by employing the  Uehling-approximation for 
the loop \cite{pea96}. Calculation of the higher-order contribution to 
this diagram are presently in progress. 
3. Until now, the two-loop diagram VPVPb as well as 
the self energy corrected one-loop vacuum polarization contribution 
VPVPc (Fig. \ref{f:1}) where calculated only to lowest order in $(Z\alpha)$ 
utilizing the K\"all\'{e}n-Sabry polarization function 
\cite{kas55,bas88,sgs93}.

Aiming towards the completion of the exact evaluation of all 
QED-radiative corrections of order $\alpha^2$, we shall present a
calculation scheme which allows the higher-order $(Z\alpha)$-contribution 
of the two-loop vacuum polarization correction VPVPb to be determined.
Recognizing that the diagram we wish to calculate is part of the complete
''dressed'' one-loop vacuum polarization allows us to reduce the problem 
to the evaluation of effective one-loop corrections where the 
renormalization procedure is known \cite{sam88,pea93}. 

Section II contains a general discussion of ''dressed'' electron
propagators in arbitrary, classical external fields and of the induced  
vacuum polarization. In section III we specify the formulae to the     
situation in hydrogen-like ions. The subtraction scheme for deducing the 
two-loop correction from various effective one-loop contributions
will be introduced. Section IV briefly reviews the renormalization 
procedure. An expression for the renormalized energy shift due to 
the higher-order (in $(Z\alpha)$) part is derived. 
This consists of two terms which require a different numerical treatment.
In order to compare the effect due to higher orders in $(Z\alpha)$ we also
present results for the two-loop correction in Uehling approximation, 
which will be derived in section V.
In section VI we will calculate the effect of the higher-order contribution
to the $1$S-Lamb shift in hydrogen-like Lead and Uranium. 

Throughout this paper, units will be used where $\hbar = m_0 = 1$ and 
$e^2 = \alpha$.

\section{Dressed electron propagators and one-loop vacuum polarization}
We start with a brief discussion of the concept of ''dressed''
electron propagators and of the corresponding ''dressed'' one-loop 
vacuum polarization (VP). Some general formulae will be derived, which 
will be employed in the next section.

We shall adopt the term ''dressed'' electron (positron) line
$\psi$ for an electron (positron) moving in an arbitrary, external 
electromagnetic field $A_\mu$. The wave function that accounts for the 
interaction with this external field is a solution of the Dirac equation: 
\begin{eqnarray}
\label{eq:1}
\left[{\rm i}\partial\!\!\!/_x -e A\!\!\!/(x) -m\right]\psi (x) = 0\quad .
\end{eqnarray}
Choosing free electron lines as a reference, we shall refer to the state
$\psi$ as ''$A$-dressed'' electron line. Similarly, an electron interacting 
with the external Coulomb-potential $V^{{\rm C}}=e A^{{\rm C}}_0$ generated 
by the (bare) nuclear charge density distribution may be called 
a ''Coulomb-dressed'' electron. 
In general situations it is appropriate to divide the
total external field $A_\mu$ into two parts $A^{{\rm e}}_\mu$ and 
$\widetilde{A}_\mu$, i.e.:
\begin{eqnarray}
\label{eq:2}
A_\mu (x) = A^{\rm e}_\mu (x) + \widetilde{A}_\mu (x) \quad ,
\end{eqnarray}
where e.g. the second term $\widetilde{A}_\mu$ may be considered a 
perturbation. Accordingly, considering electron states $\phi$ in the 
external field $A^{\rm e}_\mu$ as unperturbed states, we may then 
call these states $\psi$ ''$\widetilde{A}$-dressed''.
The propagators ${\cal S}^{{\rm A}}_{{\rm F}}(x,x')$ 
and $ S^{\rm e}_{{\rm F}}(x,x')$ describing electrons in the 
external field $A_\mu$ respectively $A^{e}_\mu$ are defined by
\begin{eqnarray}
\label{eq:3}
\left[{\rm i}\partial\!\!\!/_x -e A\!\!\!/(x) -m\right]
{\cal S}^{{\rm A}}_{{\rm F}}(x,x') &=& \delta (x-x')\quad ,\nonumber \\
\left[{\rm i}\partial\!\!\!/_x -e A\!\!\!/^{\rm e}(x) -m\right]
S^{\rm e}_{{\rm F}}(x,x') &=& \delta (x-x')\quad .
\end{eqnarray}
Note that they also satisfy the equations
\begin{eqnarray}
\label{eq:4}
{\cal S}^{{\rm A}}_{{\rm F}}(x,x')
\left[{\rm i}\partial\!\!\!/_{x'} + e A\!\!\!/(x') + m\right]
 &=& - \delta (x-x')\quad ,\nonumber \\
S^{\rm e}_{{\rm F}}(x,x') 
\left[{\rm i}\partial\!\!\!/_{x'} +e A\!\!\!/^{\rm e}(x') +m\right]
&=& -\delta (x-x')\quad ,
\end{eqnarray}
where the (adjoint) Dirac-operators are acting to the left.
In analogy to the Dyson equation defining dressed propagators in terms
of improper self-energy insertions etc. we postulate an equation of the 
form
\begin{eqnarray}
\label{eq:5}
{\cal S}^{{\rm A}}_{{\rm F}}(x,x') &=& S^{\rm e}_{{\rm F}}(x,x') + 
\int {\rm d}^4x_1\,{\rm d}^4x_2\,S^{\rm e}_{{\rm F}}(x,x_1) 
\widetilde{{\cal K}}(x_1,x_2) S^{\rm e}_{{\rm F}}(x_2,x') \quad ,
\end{eqnarray}
with a kernel $\widetilde{{\cal K}}$. With the aid of
equations (\ref{eq:3}) and (\ref{eq:4}) we can solve for the kernel:
\begin{eqnarray}
\label{eq:6}
\widetilde{{\cal K}}(x,x') &=& e\widetilde{A}\!\!\!/(x)\,\delta (x-x')
+ e\widetilde{A}\!\!\!/(x)\,{\cal S}^{{\rm A}}_{{\rm F}}(x,x') 
e\widetilde{A}\!\!\!/(x')\quad .
\end{eqnarray}
Insertion of Eq. (\ref{eq:6}) into (\ref{eq:5}) leads to an
equation for the ''$\widetilde{A}$-dressed''
propagator ${\cal S}^{{\rm A}}_{{\rm F}}$ (taking $S^{\rm e}_{{\rm F}}$ as 
the reference) which can be solved iteratively. 
The exact propagator ${\cal S}^{{\rm A}}_{{\rm F}}$
appears as the sum of the unperturbed external field propagator
$S^{\rm e}_{{\rm F}}$, a part describing a single-scattering with the
additional external field $\widetilde{A}_\mu$ and a higher-order part which
accounts for multiple-scattering contributions.
Given a representation for ${\cal S}^{{\rm A}}_{{\rm F}}$ a formal 
expression for the corresponding one-loop vacuum-polarization current
induced by the total external field $A_\mu$ can be derived. 
For later purposes we will already specialize to the case of static external 
fields. Since the propagators are homogeneous in time, one obtains:
\begin{eqnarray}
\label{eq:8}
{\cal J}^{{\rm A}\mu}(\bbox r) = &&
{\rm i}e \int \frac{{\rm d}E}{2\pi}\,
{\rm Tr}\left[\gamma^\mu{\cal S}^{{\rm A}}_{{\rm F}}(\bbox r,\bbox r,E)\right] 
\nonumber \\
&&\hspace{-1.5cm}=\, {\rm i}e \int \frac{{\rm d}E}{2\pi}\,
\left\{ {\rm Tr}\left[\gamma^\mu S^{{\rm e}}_{{\rm F}}(\bbox r,\bbox r,E)\right] 
+ \int {\rm d}^3r_1\,
{\rm Tr}\left[\gamma^\mu S^{{\rm e}}_{{\rm F}}(\bbox r,\bbox r_1,E)\,
e\widetilde{A}\!\!\!/(\bbox r_1)\,
S^{{\rm e}}_{{\rm F}}(\bbox r_1,\bbox r,E)\right]
\right. \nonumber \\
&& \hspace{-1.5cm}\left. + \int {\rm d}^3r_1\,{\rm d}^3r_2\,
{\rm Tr}\left[\gamma^\mu S^{{\rm e}}_{{\rm F}}(\bbox r,\bbox r_1,E)\, 
e\widetilde{A}\!\!\!/(\bbox r_1)\,
{\cal S}^{{\rm A}}_{{\rm F}}(\bbox r_1,\bbox r_2,E)\,
e\widetilde{A}\!\!\!/(\bbox r_2)\,
S^{{\rm e}}_{{\rm F}}(\bbox r_2,\bbox r,E)\right]\right\} \quad .
\end{eqnarray}
Putting aside questions about renormalization for a moment, this formally 
exact equation implies that the one-loop vacuum polarization
${\cal J}^{{\rm A}\mu}$ induced by the total field $A_\mu$ is given
as a sum of three terms: a part 
induced by the external field $A^{{\rm e}}_\mu$, a single-interaction
contribution and a third part taking into account multiple interactions 
with the additional external field $\widetilde{A}_\mu$. 

The representation of the propagator 
${\cal S}^{{\rm A}}_{{\rm F}}$ and of the induced vacuum 
polarization ${\cal J}^{{\rm A}\mu}$ derived above are not unique. 
The reason for this is provided by the fact that the decomposition of the
total external field (\ref{eq:2}) is completely arbitrary. In particular,
we could have chosen the free-field configuration as unperturbed reference.
Consequently, the external field propagator $S^{{\rm e}}_{{\rm F}}$ in the 
defining equation (\ref{eq:5}) has to be replaced by the free Feynman 
propagator $S^{{\rm 0}}_{{\rm F}}$ leading to a similar kernel (\ref{eq:6})
which will contain the total external field $A_\mu$. 
In this case Eq. (\ref{eq:8}) takes the form 
\begin{eqnarray}
\label{eq:9}
{\cal J}^{{\rm A}\mu}(\bbox r) 
=&& {\rm i}e \int \frac{{\rm d}E}{2\pi}\,\left\{ \int {\rm d}^3r_1\,
{\rm Tr}\left[\gamma^\mu S^{{\rm 0}}_{{\rm F}}(\bbox r-\bbox r_1,E)\,
e A\!\!\!/(\bbox r_1)\,
S^{{\rm 0}}_{{\rm F}}(\bbox r_1-\bbox r,E)\right]
\right. \nonumber \\
&& \hspace{-1.5cm}\left. + \int {\rm d}^3r_1\,{\rm d}^3r_2\,
{\rm Tr}\left[\gamma^\mu S^{{\rm 0}}_{{\rm F}}(\bbox r-\bbox r_1,E)\, 
e A\!\!\!/(\bbox r_1)\,
{\cal S}^{{\rm A}}_{{\rm F}}(\bbox r_1,\bbox r_2,E)\,
e A\!\!\!/(\bbox r_2)\,
S^{{\rm 0}}_{{\rm F}}(\bbox r_2-\bbox r,E)\right]\right\} \, .
\nonumber \\
\end{eqnarray}
The free closed-loop contribution vanishes in accordance with the 
Furry-theorem. 
The induced vacuum polarization itself gives rise to a modification
${\cal A}_\mu$ of the total external field:
\begin{eqnarray}
\label{eq:10}
{\cal A}_\mu (\bbox r) &=& \int {\rm d}^3r'\,
D_{\mu\nu} (\bbox r - \bbox r',0)\,{\cal J}^{{\rm A}\nu}(\bbox r')
\quad .
\end{eqnarray}
The free photon propagator is given by (in Feynman gauge)
\begin{eqnarray}
\label{eq:10a}
D_{\mu\nu} (\bbox r - \bbox r',0) &=& g_{\mu\nu}\,D(\bbox r - \bbox r',0) = 
- g_{\mu\nu} \int \,\frac{{\rm d}^3k}{(2\pi)^3}\,
e^{-{\rm i} \bbox k\cdot (\bbox r - \bbox r')}\,
\frac{1}{- \bbox k^2 + {\rm i}\varepsilon}\nonumber \\
&=& g_{\mu\nu}\,\frac{1}{\pi} \int_0^\infty \,{\rm d}k\,\sum_{\ell,m}\,4\pi\,
Y_{\ell m}(\hat{r})\,Y^\ast_{\ell m}(\hat{r}')\,
{\rm j}_\ell (kr)\,{\rm j}_\ell (kr')
\quad .
\end{eqnarray}
The last line of the expression above specifies the partial wave 
decomposition of the photon propagator.
A spherically symmetric, static external potential $V = e A_0$ induces 
only a static vacuum polarization charge density ${\cal J}^{{\rm A} 0}$  
which will also be spherically symmetric. 
It gives rise to an effective one-loop VP-potential:
\begin{eqnarray}
\label{eq:11}
{\cal V}^{{\rm A}}(r) &=& \frac{e}{\pi} \int \,{\rm d}k\,
{\rm j}_0 (kr)\,\int \,{\rm d}r'\,r'^2\,{\rm j}_0 (kr')\,
{\cal J}^{{\rm A} 0}(r') \quad .
\end{eqnarray}

\section{Subtraction scheme}
\begin{figure}
\centerline{\mbox{\epsfxsize=10cm\epsffile{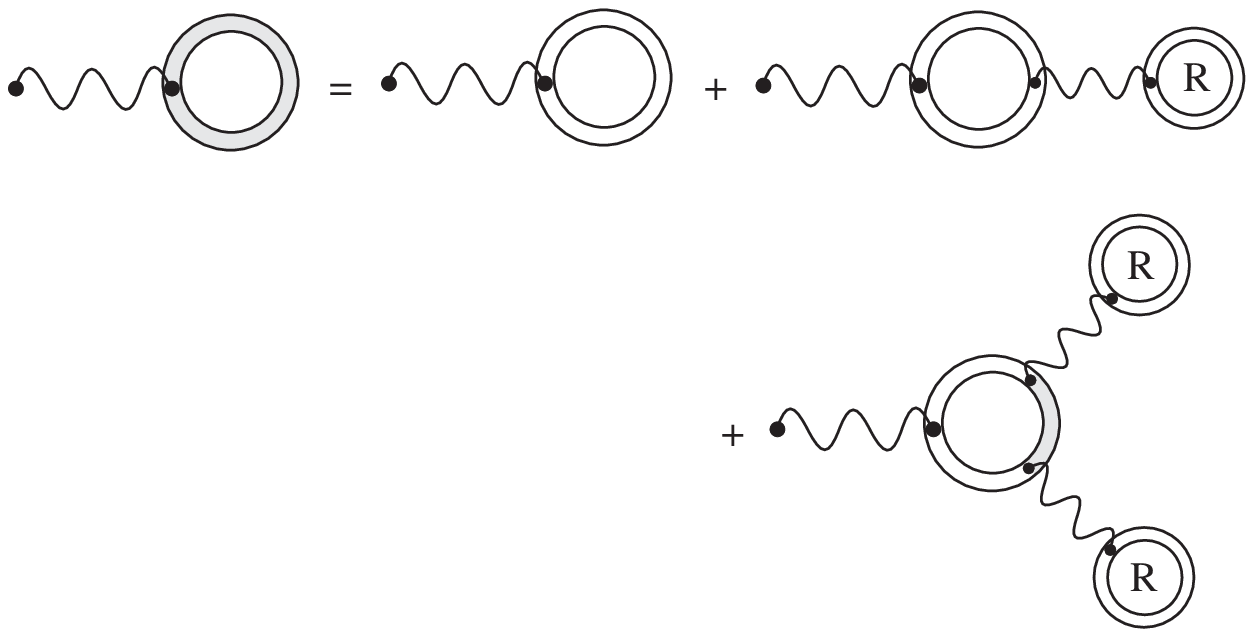}}} 
\caption{\label{f:2} VP-dressed one-loop vacuum polarization potential
(indicated by shadowed lines). The ''R'' inside of the external VP-loops
refers to the {\em renormalized} one-loop VP-potential 
$V^{{\rm VP}}_{{\rm ren}}$ induced by the external Coulomb field of the 
nucleus, which appears as additional external potential.}
\end{figure}

The general considerations of the previous section may have already 
anticipated how we are going to deduce the two-loop vacuum polarization 
correction. The basic idea is to derive this contribution from an 
effective one-loop vacuum polarization ${\cal V}^{{\rm A}}$, 
which is dressed with the renormalized first-order vacuum polarization 
potential $V^{{\rm VP}}_{{\rm ren}}$ induced by the external Coulomb field 
$V^{{\rm C}}$ of the nucleus. At first we need
to specialize Eq. (\ref{eq:8}) to (\ref{eq:11}) to the situation of
bound-state QED. In the presence of a static, spherically symmetric
nuclear charge density, we specify the total external field (\ref{eq:2})
as the sum of the (bare) external Coulomb potential $V^{{\rm C}}$ and
the renormalized, first-order vacuum polarization potential
$V^{{\rm VP}}_{{\rm ren}}$:
\begin{eqnarray}
\label{eq:12}
V(r) &=& e A_0(r) = V^{{\rm C}}(r) + V^{{\rm VP}}_{{\rm ren}}(r)\quad .
\end{eqnarray}
The one-loop potential is obtained from
\begin{eqnarray}
\label{eq:13}
V^{{\rm VP}}_{{\rm ren}}(r) &=& e \widetilde{A}_0(r) = 
{\rm i} \alpha \int {\rm d}^3r'\,D(\bbox r - \bbox r',0)\,
\left\{ \int \frac{{\rm d}E}{2\pi}\,
{\rm Tr}\left[\gamma^0 S^{{\rm C}}_{{\rm F}}(\bbox r',\bbox r',E)\right]
\right\}_{{\rm ren}}
\end{eqnarray}
after renormalization. $S^{{\rm C}}_{{\rm F}}$ denotes the electron
propagator in the external Coulomb field. 
In view of Eqs. (\ref{eq:8}) and (\ref{eq:11}) the VP-dressed one-loop vacuum 
polarization potential formally reads
\begin{eqnarray}
\label{eq:14}
{\cal V}(r) &=& {\rm i} \alpha \int {\rm d}^3r'\,
D(\bbox r - \bbox r',0)\,\int \frac{{\rm d}E}{2\pi}\,
{\rm Tr}\left[\gamma^0 {\cal S}^{{\rm V}}_{{\rm F}}(\bbox r',\bbox r',E)\right]
\nonumber \\
&=& {\rm i} \alpha \int {\rm d}^3r'\,D(\bbox r - \bbox r',0)\,
\int \frac{{\rm d}E}{2\pi}\,\left\{ \phantom{\frac{1}{2}}
{\rm Tr}\left[\gamma^0 S^{{\rm C}}_{{\rm F}}(\bbox r',\bbox r',E)\right] 
\right. \nonumber \\
&& \left. + \int {\rm d}^3r_1\,
{\rm Tr}\left[\gamma^0 S^{{\rm C}}_{{\rm F}}(\bbox r',\bbox r_1,E)\,
\gamma^0 V^{{\rm VP}}_{{\rm ren}}(r_1)\,
S^{{\rm C}}_{{\rm F}}(\bbox r_1,\bbox r',E)\right]\right.  \\
&& \left.\hspace{-1.5cm}+ \int {\rm d}^3r_1\,{\rm d}^3r_2\,
{\rm Tr}\left[\gamma^0 S^{{\rm C}}_{{\rm F}}(\bbox r',\bbox r_1,E)\,
\gamma^0 V^{{\rm VP}}_{{\rm ren}}(r_1)\,
{\cal S}^{{\rm V}}_{{\rm F}}(\bbox r_1,\bbox r_2,E)
\gamma^0 V^{{\rm VP}}_{{\rm ren}}(r_2)\,
S^{{\rm C}}_{{\rm F}}(\bbox r_2,\bbox r',E)\right]
\right\} \quad .  \nonumber
\end{eqnarray}
A graphical representation of the VP-dressed one-loop potential is depicted 
in Fig. \ref{f:2}.
The two-loop vacuum polarization correction we are interested in appears
as the second term of the decomposition (\ref{eq:14}) together with the
ordinary one-loop vacuum polarization (first term). The third term accounts
for all multiple interaction terms and is at least of order 
$\alpha^4(Z\alpha)^3$, since the additional potential
$V^{{\rm VP}}_{{\rm ren}}$ contributes with a leading order
$\alpha (Z\alpha)$ (Uehling-term). 
Thus, it is legitimate to neglect this higher-order part. Consequently,
we identify as the renormalized two-loop potential
\begin{eqnarray}
\label{eq:15}
{\cal U}^{{\rm VPVPb}}_{{\rm ren}}(r) &\simeq& 
{\rm i} \alpha \int {\rm d}^3r'\,D(\bbox r - \bbox r',0)\,
\left\{\left(\int \frac{{\rm d}E}{2\pi}\,
{\rm Tr}\left[\gamma^0 {\cal S}^{{\rm V}}_{{\rm F}}(\bbox r',\bbox r',E)
\right]\right) \right.\nonumber \\
&& \hspace{4.1cm} \left. - \left(\int \frac{{\rm d}E}{2\pi}\,
{\rm Tr}\left[\gamma^0 S^{{\rm C}}_{{\rm F}}(\bbox r',\bbox r',E)\right]
\right)\,\right\}_{{\rm ren}}\quad , 
\end{eqnarray}
after appropriate renormalizations have been applied to the right-hand
side of the above equation. Note, however, that the formal expression 
for the effective one-loop vacuum polarization density (first term in the 
curly brackets) already contains the renormalized one-loop potential 
$V^{{\rm VP}}_{{\rm ren}}$,
assuming that the renormalization of the exterior VP-loops can be performed
separately. This issue will be adressed in the next section.
We are now in the position to write down the energy shift of a bound 
electron state $\phi_A$ due to the exact two-loop correction:
\begin{eqnarray}
\label{eq:16}
\Delta E^{{\rm VPVPb}}_A = 
\langle\phi_A|{\cal U}^{{\rm VPVPb}}_{{\rm ren}}|\phi_A\rangle 
\quad .
\end{eqnarray}
In order to deduce exclusively the contribution 
$\Delta E^{{\rm VPVPb}}_A({\rm h.o.})$
arising from higher orders in $(Z\alpha)$, one has to 
subtract the corresponding two-loop contribution 
evaluated in the Uehling-approximation.

\section{Renormalization}
\begin{figure}
\centerline{\mbox{\epsfxsize=10cm\epsffile{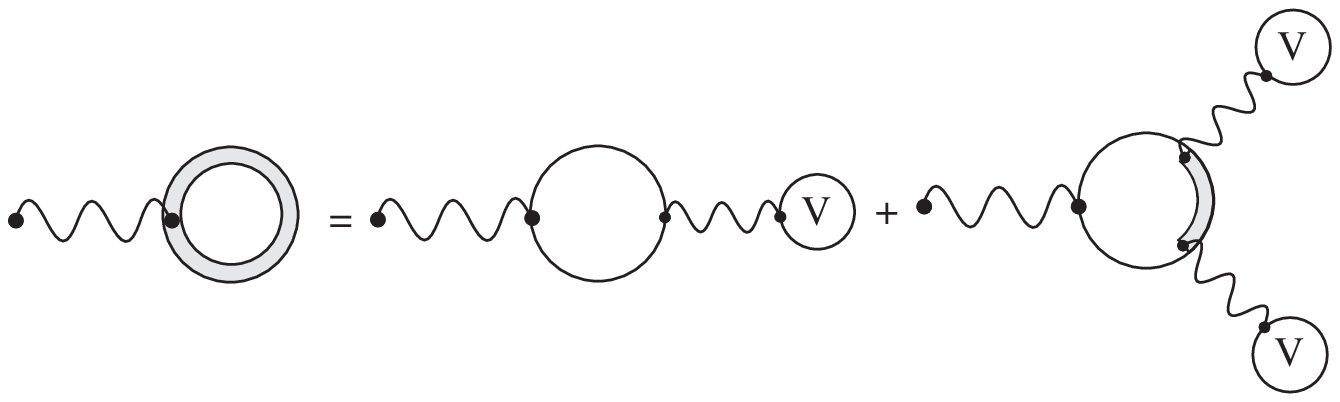}}} 
\caption{\label{f:3} $V$-dressed one-loop vacuum polarization potential
(indicated by shadowed lines) according to the decomopsition 
(\protect\ref{eq:8}). ${\bf \bigcirc \! \!\!\! {\rm v}}$ symbolizes 
interactions with the total (renormalized) external potential Eq. 
(\protect\ref{eq:12}).}
\end{figure}

\begin{figure}
\centerline{\mbox{\epsfxsize=15cm\epsffile{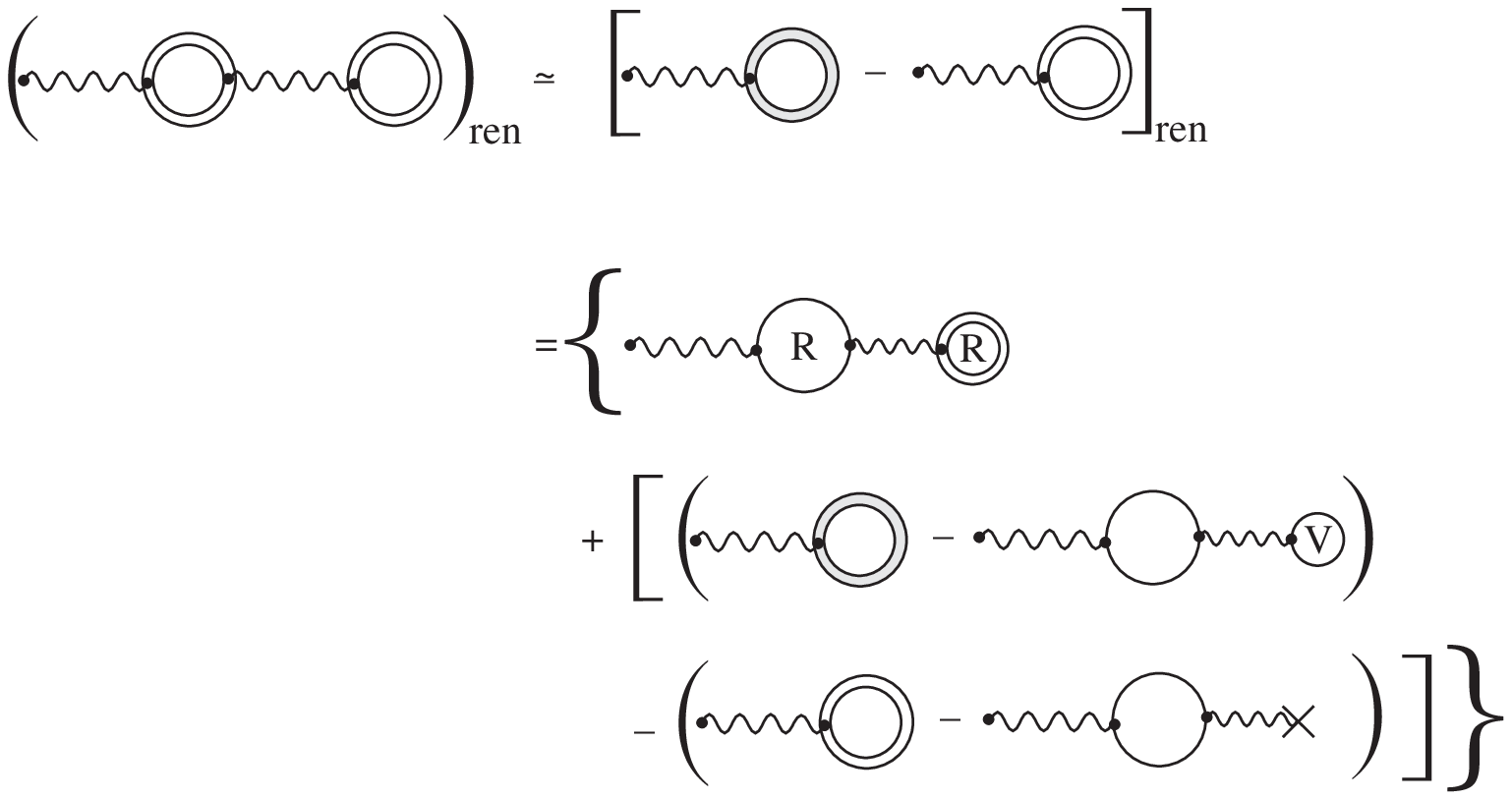}}}
\caption{\label{f:4} Diagrammatic representation the subtraction scheme
for deducing the renormalized two-loop vacuum polarization potential
according to Eqs. (\protect\ref{eq:22}) -- (\protect\ref{eq:24}). The first 
term in curly brackets stands for the part 
${\cal U}^{{\rm VPVPb}}_{{\cal F}1}$ and the second term in square
brackets symbolizes the
Wichmann-Kroll-type potential ${\cal U}^{{\rm VPVPb}}_{{\cal F}2}$.}
\end{figure}

Taking the above into consideration, the problem
of renormalization of the exact two-loop potential 
\begin{eqnarray}
\label{eq:17}
{\cal U}^{{\rm VPVPb}}(r) &=& {\rm i} \alpha \int {\rm d}^3r'\,
D(\bbox r - \bbox r',0) \int {\rm d}^3r_1\,
\int \frac{{\rm d}E}{2\pi}\,
{\rm Tr}\left[\gamma^0 S^{{\rm C}}_{{\rm F}}(\bbox r',\bbox r_1,E)\,
\gamma^0\,S^{{\rm C}}_{{\rm F}}(\bbox r_1,\bbox r',E)\right] \nonumber \\ 
&& \times \int {\rm d}^3r_2\, D(\bbox r_1 - \bbox r_2,0)
\int \frac{{\rm d}E'}{2\pi}\,
{\rm Tr}\left[\gamma^0 S^{{\rm C}}_{{\rm F}}(\bbox r_2,\bbox r_2,E')\right]
\end{eqnarray}
reduces to the problem of renormalization of the one-loop vacuum
polarization. This is sugguested by Eqs. (\ref{eq:14}) and (\ref{eq:15}). 
It relies on the fact, that the external VP-loop is properly taken into 
account in terms of a VP-dressed effective one-loop, where the external 
VP-loop may be replaced by the renormalized first-order potential 
$V^{{\rm VP}}_{{\rm ren}}$. This is supported by the notion that the 
electron (positron) experiences the effective nuclear charge modified by the
induced vacuum polarization cloud. Procedures for renormalization of
one-loop potentials involved in Eq. (\ref{eq:15}) are well known from 
the evaluation of the energy shift due to the first-order vacuum polarization
in external Coulomb fields. We adopt the partial wave decomposition 
approach together with the subtraction scheme
developed in \cite{sam88,pea93} and apply it to Eq. (\ref{eq:15}).
In order to perform similar steps as in the case of the ordinary 
Coulomb-dressed VP-loop, we employ the equivalent representation 
according to Eq. (\ref{eq:9})
\begin{eqnarray}
\label{eq:18}
{\cal V}(r) &=& {\rm i} \alpha \int {\rm d}^3r'\,D(\bbox r - \bbox r',0)\,
\int \frac{{\rm d}E}{2\pi}\,\left\{ 
\int {\rm d}^3r_1\,
{\rm Tr}\left[\gamma^0 S^0_{{\rm F}}(\bbox r'-\bbox r_1,E)\,
\gamma^0 \,V(r_1)\,
S^0_{{\rm F}}(\bbox r_1 - \bbox r',E)\right]\right. \nonumber \\
&& \left.\hspace{-1.5cm}+ \int {\rm d}^3r_1\,{\rm d}^3r_2\,
{\rm Tr}\left[\gamma^0 S^0_{{\rm F}}(\bbox r'-\bbox r_1,E)\,
\gamma^0 \,V(r_1)\,
{\cal S}^{{\rm V}}_{{\rm F}}(\bbox r_1,\bbox r_2,E)
\gamma^0 \,V(r_2)\,
S^0_{{\rm F}}(\bbox r_2 -\bbox r',E)\right]
\right\} \quad ,  
\end{eqnarray}
where $V$ denotes the total external potential Eq. (\ref{eq:12}).  
Fig. \ref{f:3} shows the diagrammatic representation of Eq. (\ref{eq:18}).
Further evaluation requires a partial wave decomposion of all the
propagators    
\begin{eqnarray}
\label{eq:19}
{\cal S}^{{\rm V}}_{{\rm F}}(\bbox r,\bbox r',E) &=& \sum_{n\kappa\mu}\,
\frac{\psi_{n\kappa\mu}(\bbox r) \overline{\psi}_{n\kappa\mu}(\bbox r')}
{E - {\cal E}_{n,\kappa,\mu}(1-{\rm i}\eta)} \quad , \nonumber \\
S^{{e}}_{{\rm F}}(\bbox r,\bbox r',E) &=& \sum_{n\kappa\mu}\,
\frac{\phi_{n\kappa\mu}(\bbox r) \overline{\phi}_{n\kappa\mu}(\bbox r')}
{E - E_{n,\kappa}(1-{\rm i}\eta)} \quad , \nonumber \\
S^0_{{\rm F}}(\bbox r-\bbox r',E) &=& \sum_{p,\kappa,\mu}\,
\frac{\varphi_{p\kappa\mu}(\bbox r) \overline{\varphi}_{p\kappa\mu}(\bbox r')}
{E - \varepsilon_{p,\kappa}(1-{\rm i}\eta)} \quad .
\end{eqnarray}
together with Eq. (\ref{eq:10a}) for the photon propagator.  
According to the renormalization prescription developed in \cite{pea93} 
the renormalized one-loop potential ${\cal V}_{{\rm ren}}$ is obtained as
the sum of the finite ''Wichmann-Kroll''-type contribution 
${\cal V}_{{\cal F}2}$ and of the renormalized ''Uehling''-type potential   
${\cal V}_{{\cal F}1}$. 
We are lead to the following expressions: 
\begin{eqnarray}
\label{eq:20}
{\cal V}_{{\cal F}2}(r) &=& - \frac{\alpha}{\pi}\int {\rm d}k\,{\rm j}_0(kr)\,
\sum_{|\kappa |=1}^{|\kappa_{{\rm max}}|}\,
\left\{ \sum_{n} \,{\rm sign}({\cal E}_{n,\kappa})\,
\langle\psi_{n\kappa}|\,{\rm j}_0(kr')\,|\psi_{n\kappa}\rangle 
\right.\nonumber \\
& & \left. \hspace{4.5cm} - \,4 \sum_p^+ \sum_{p'}^-\,
\frac{\langle\varphi_{p\kappa}|\,{\rm j}_0(kr')\,|\varphi_{p'\kappa}
\rangle\langle\varphi_{p'\kappa}|\,V\,|\varphi_{p\kappa}\rangle}
{\varepsilon_{p,\kappa}-\varepsilon_{p',\kappa}}\right\}\quad ,
\nonumber \\
{\cal V}_{{\cal F}1}(r) &=& \int_0^\infty\,{\rm d}r'\,4\pi r'^2\,
\left[{\rm e}\rho^{{\rm V}}(r')\right]\,f(r,r')\quad ,
\end{eqnarray}
together with the radial kernel
\begin{eqnarray}
\label{eq:21}
f(r,r') &=& - \frac{2\alpha}{3\pi} \int_1^\infty {\rm d}\xi \,
\sqrt{1-\frac{1}{\xi^2}} \left(1+\frac{1}{2\xi^2}\right)\,
\left[\Theta(r-r')\,
\frac{e^{-2r \xi}}{r\xi}\,
\frac{{\rm sinh}(2r' \xi)}{2r' \xi} 
\right.\nonumber \\
&&\left. \hspace{6.3cm} + \Theta(r'-r)\,
\frac{e^{-2r' \xi}}{r'\xi}\,
\frac{{\rm sinh}(2r \xi)}{2r \xi}\right]
\nonumber \\[0.3cm]
&=& - \frac{\alpha}{3\pi r'}\,\left(\frac{1}{2r }\right)\,
\left[ \chi_2(2|r-r'| ) - \chi_2(2(r+r') )\right]
\quad ,\nonumber \\[0.3cm]
\chi_n (z) &=& \int_1^\infty {\rm d}\xi\,\sqrt{1-\frac{1}{\xi^2}}\,
\left(1+\frac{1}{2\xi^2}\right) \frac{e^{-z\xi}}{\xi^n}\quad .
\end{eqnarray}
$\rho^{{\rm V}}$ denotes the sum of the (bare) nuclear charge density
$\rho_{{\rm nuc}}$ and of the renormalized first-order vacuum polarization
charge density $\rho^{{\rm VP}}_{{\rm ren}}$ induced by the Coulomb field 
of the nucleus. 
Eqs. (\ref{eq:20}) and (\ref{eq:21}) are analogous to
expressions one has to deal with when deriving the renormalized, exact
one-loop potential (\ref{eq:13}). Performing now the subtraction as
implied by Eq. (\ref{eq:15}), we identify the renormalized two-loop
vacuum polarization potential:
\begin{eqnarray}
\label{eq:22}
{\cal U}^{{\rm VPVPb}}_{{\rm ren}}(r) &=& 
{\cal U}^{{\rm VPVPb}}_{{\cal F}1}(r) + {\cal U}^{{\rm VPVPb}}_{{\cal F}2}(r) 
\quad ,  \\[0.3cm]
\label{eq:23}
{\cal U}^{{\rm VPVPb}}_{{\cal F}1}(r) &=& \int_0^\infty\,{\rm d}r'\,4\pi r'^2\,
\left[{\rm e}\rho^{{\rm VP}}_{{\rm ren}}(r')\right]\,f(r,r')\quad ,\\
\label{eq:24}
{\cal U}^{{\rm VPVPb}}_{{\cal F}2}(r) &\simeq& 
- \frac{\alpha}{\pi}\int {\rm d}k\,{\rm j}_0(kr)\,
\sum_{|\kappa |=1}^{|\kappa_{{\rm max}}|}\,\left\{
\left[\sum_{n} \,{\rm sign}({\cal E}_{n,\kappa})\,
\langle\psi_{n\kappa}|\,{\rm j}_0(kr')\,|\psi_{n\kappa}\rangle 
\right.\right.\nonumber \\
& & \left. \left.\hspace{0.0cm} - \,4 \sum_p^+ \sum_{p'}^-\,
\frac{\langle\varphi_{p\kappa}|\,{\rm j}_0(kr')\,|\varphi_{p'\kappa}
\rangle\langle\varphi_{p'\kappa}|\,V\,|\varphi_{p\kappa}\rangle}
{\varepsilon_{p,\kappa}-\varepsilon_{p',\kappa}}\right]
\right.\nonumber \\
&& \left.\hspace{-2.0cm} 
- \left[\sum_n \,{\rm sign}(E_{n,\kappa})\,
\langle\phi_{n\kappa}|\,{\rm j}_0(kr')\,|\phi_{n\kappa}\rangle 
- \,4 \sum_p^+ \sum_{p'}^-\,
\frac{\langle\varphi_{p\kappa}|\,{\rm j}_0(kr')\,|\varphi_{p'\kappa}
\rangle\langle\varphi_{p'\kappa}|\,V^{{\rm C}}\,|\varphi_{p\kappa}\rangle}
{\varepsilon_{p,\kappa}-\varepsilon_{p',\kappa}}\right]\right\}\quad .
\nonumber \\
\end{eqnarray}
The term ${\cal U}^{{\rm VPVPb}}_{{\cal F}1}$ denotes the finite
(renormalized) Uehling-type potential generated by the Uehling part of 
the renormalized first-order vacuum polarization charge density. The 
term ${\cal U}^{{\rm VPVPb}}_{{\cal F}2}$ summarizes the finite part of the
Wichmann-Kroll-type potential. Since the sum over partial waves terminates
at some finite $|\kappa_{{\rm max}}|$ each contribution in square
brackets becomes well defined. The procedure described in Eqs. (\ref{eq:22})
to (\ref{eq:24}) for deducing the renormalized exact two-loop vacuum
polarization potential is depicted in Fig. \ref{f:4}.

For numerical evaluations, however, the representation (\ref{eq:24}) 
of the potential ${\cal U}^{{\rm VPVPb}}_{{\cal F}2}$ is ruther cumbersome 
since it involves terms given as multiple summations 
over the free Dirac spectrum. 
A representation which is more convenient for numerical calculations can be 
introduced as follows: At first, we combine the one-potential terms in 
Eq. (\ref{eq:24}) to a one-potential term involving the renormalized
one-loop vacuum potential $V^{{\rm VP}}_{{\rm ren}}$ only. Secondly,
the resulting one-potential term is replaced by the completely 
VP-dressed (free) one-loop potential assuming that the effects due to
multiple-interaction contributions, which are at least of order 
$\alpha (\alpha (Z\alpha))^3$, are neglegible. Thus we obtain the alternative
representation:
\begin{eqnarray}
\label{eq:24b}
{\cal U}^{{\rm VPVPb}}_{{\cal F}2}(r) &\simeq& 
- \frac{\alpha}{\pi}\int {\rm d}k\,{\rm j}_0(kr)\,
\sum_{|\kappa |=1}^{|\kappa_{{\rm max}}|}\,\left\{
\sum_{n} \,{\rm sign}({\cal E}_{n,\kappa})\,
\langle\psi_{n\kappa}|\,{\rm j}_0(kr')\,|\psi_{n\kappa}\rangle 
\right.\nonumber \\
&& \left.
- \sum_n \,{\rm sign}(E_{n,\kappa})\,
\langle\phi_{n\kappa}|\,{\rm j}_0(kr')\,|\phi_{n\kappa}\rangle 
- \sum_n \,{\rm sign}(\widetilde{\varepsilon}_{n,\kappa})\,
\langle\widetilde{\varphi}_{n\kappa}|\,
{\rm j}_0(kr')\,|\widetilde{\varphi}_{n\kappa}\rangle \right\}\quad ,
\nonumber \\
\end{eqnarray}
where the states $\widetilde{\varphi}_{n\kappa}$ are solutions of the 
Dirac equation with the external potential $V^{{\rm VP}}_{{\rm ren}}$ and 
corresponding energy eigenvalues $\widetilde{\varepsilon}_{n,\kappa}$.

Having derived the renormalized potential 
${\cal U}^{{\rm VPVPb}}_{{\rm ren}}$, the corresponding energy shift of a 
bound electron state $\phi_A$ in the external Coulomb field can be 
evaluated according to (\ref{eq:16}). 

\section{Two-loop contribution in Uehling approximation}

\begin{figure}
\centerline{\mbox{\epsfxsize=15cm\epsffile{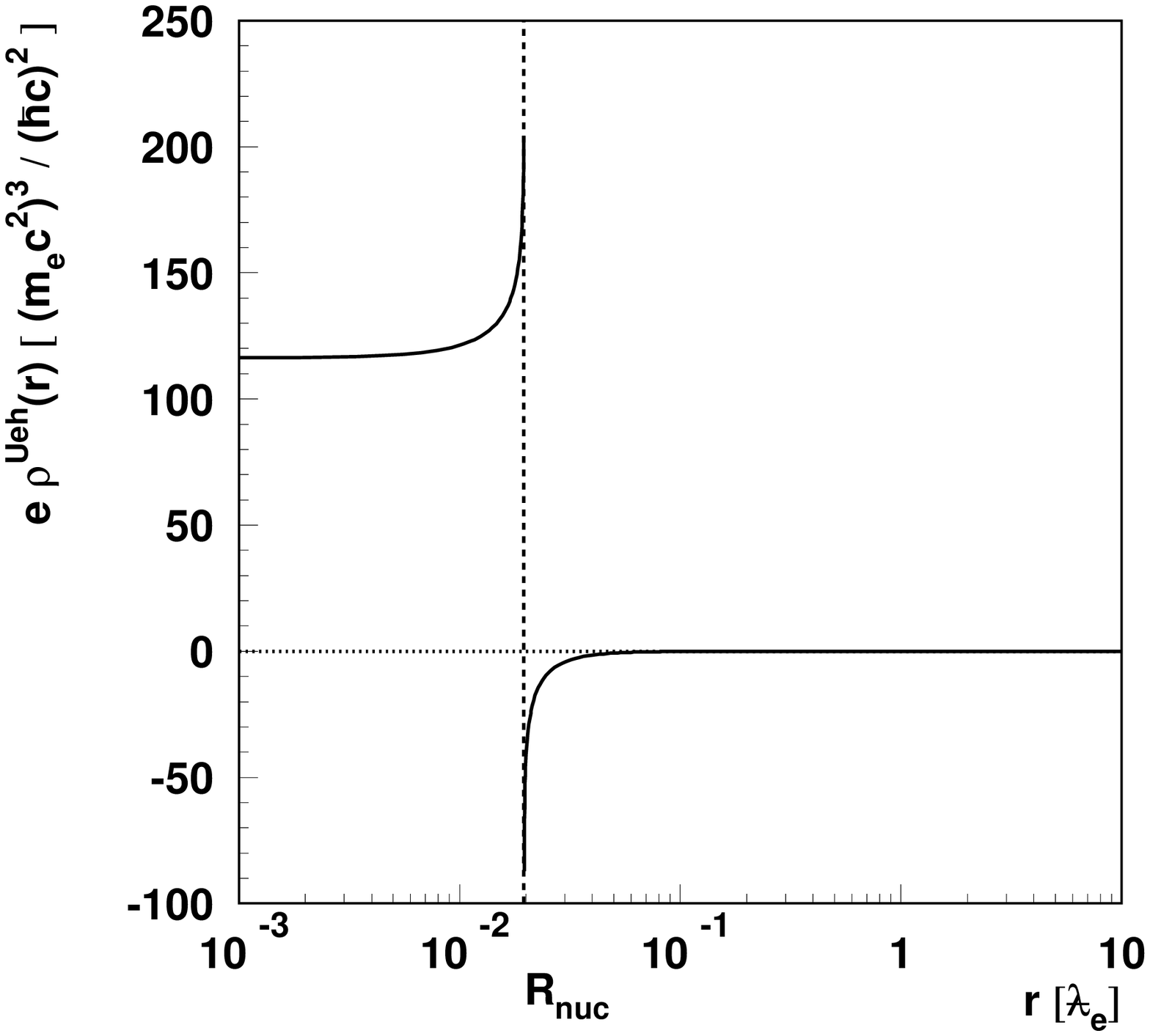}}} 
\caption{ \label{f:5} Uehling-vacuum polarization charge density
$e\rho^{{\rm Ueh}}_{{\rm ren}}$ induced by the extended external charge of 
a Uranium nucleus (uniform sphere model) as a function of the radial
distance $r$. Natural units are used.}
\end{figure}

Being interested primarily in the contribution to the energy shift due 
to higher orders in $(Z\alpha)$, we should subtract the 
Uehling-in-Uehling contribution, where the exact vacuum polarization loops 
are replaced by free fermion loops. 
Taking a uniform sphere model for the nuclear charge distribution, i.e. 
$\rho_{{\rm nuc}} = (3Ze/4\pi R_0^3)\,\Theta (R_0-r)$, the latter reads:
\begin{eqnarray}
\label{eq:ueh1}
{\cal U}^{{\rm VPVPb}}_{{\cal F}1,{\rm Ueh}}(r) &=& 
\int_0^\infty\,{\rm d}r'\,4\pi r'^2\,
\left[{\rm e}\rho^{{\rm Ueh}}_{{\rm ren}}(r')\right]\,f(r,r')\quad ,
\nonumber \\[0.2cm]
\left[{\rm e}\rho^{{\rm Ueh}}_{{\rm ren}}(r)\right] &=& 
\left(\frac{3Z\alpha}{4\pi R_0^3}\right)\,\frac{2\alpha}{3\pi}\,
\frac{R_0}{r} \int_1^\infty {\rm d}\xi\,\sqrt{1-\frac{1}{\xi^2} }
\left(1+\frac{1}{2\xi^2}\right) \frac{1}{\xi} \nonumber \\[0.2cm]
&& \hspace{1.5cm} \times \left\{\Theta (R_0-r)\left(1+\frac{1}{2R_0 \xi}\right)
{\rm sinh}(2r \xi)\,e^{-2R_0 \xi}
\right. \nonumber \\[0.2cm]
&& \left. \hspace{1.5cm}- \Theta (r-R_0) \left[{\rm cosh}(2R_0 \xi) -
\frac{{\rm sinh}(2R_0 \xi)}{2R_0 \xi}\right]\,
e^{-2r \xi}
\right\}\quad .
\end{eqnarray}
The expression for $\rho^{{\rm Ueh}}_{{\rm ren}}$ may be cast into the more
familiar form:
\begin{eqnarray}
\label{eq:ueh2}
\left[{\rm e}\rho^{{\rm Ueh}}_{{\rm ren}}(r)\right] &=& 
\left(\frac{Z\alpha}{4\pi R_0^2}\right)\,\frac{\alpha}{\pi r}\,
\left\{ {\rm sign}(R_0-r)\,\chi_1(2|R_0-r| ) - 
\chi_1(2(R_0+r) )\right.\nonumber \\
&&\left.\hspace{2cm}+\frac{1}{2R_0 }\left[
\chi_2(2|R_0-r| ) - \chi_2(2(R_0+r) )\right]\right\}
\quad .
\end{eqnarray}
This vacuum polarization charge density is plotted in Fig. \ref{f:5}. It is  
easily verified that the induced Uehling density (\ref{eq:ueh2}) possesses
a logarithmic singularity at the nuclear radius, which originates from the
first term in the curly brackets. Integrating the 
Uehling charge density over the interior of the nucleus we obtain the finite
induced vacuum charge 
\begin{eqnarray}
\label{eq:qvac}
e Q^{{\rm int}}(R_0) &=& \int_0^{R_0} {\rm d}r'\,4\pi r'^2\,
\left[e\rho_{{\rm ren}}^{{\rm Ueh}}(r')\right] \nonumber \\
&=& \frac{Z\alpha}{2R_0}\,\frac{\alpha}{\pi}\,\left[\chi_2(0) + \chi_2(4R_0)
+ \frac{\chi_3(4R_0)}{R_0}+\frac{\chi_4(4R_0)-\chi_4(0)}{(2R_0)^2}\right]
\nonumber \\
&=& - \int_{R_0}^\infty {\rm d}r'\,4\pi r'^2\,
\left[e\rho_{{\rm ren}}^{{\rm Ueh}}(r')\right] = -\,e Q^{{\rm ext}}(R_0)
\end{eqnarray}
which is exactly cancelled by the total induced charge in the exterior 
region $Q^{{\rm ext}}$.

The corresponding energy correction to the binding energy --
we refer to it as Uehling-in-Uehling correction --  which is part of the 
K\"all\'{e}n-Sabry correction \cite{sgs93} reads:
\begin{eqnarray}
\label{eq:26b}
\Delta E^{{\rm VPVPb}}_A({\cal F}1,{\rm Ueh}) &=& \langle\phi_A|\,
{\cal U}^{{\rm VPVPb}}_{{\cal F}1,{\rm Ueh}}\,|\phi_A\rangle \quad .
\end{eqnarray}
This will be calculated separately.

\section{Evaluation}
We now turn to the evaluation of the higher-oder ($Z\alpha$)-contribution
of the two-loop vacuum polarization to the Lamb shift of strongly bound
electrons 
\begin{eqnarray}
\label{eq:26}
\Delta E^{{\rm VPVPb}}_A({\rm h.o.}) &=& \langle\phi_A|
\left({\cal U}^{{\rm VPVPb}}_{{\cal F}1,{\rm WK}} + 
{\cal U}^{{\rm VPVPb}}_{{\cal F}2}\right)|\phi_A\rangle \quad ,
\end{eqnarray}
where the Uehling-in-Uehling part is subtracted.
Thus the effect of all higher orders in $(Z\alpha)$ in the interaction
with the external Coulomb potential contributing to the exact two-loop
correction proceeds in two separate steps.
Accordingly, we only need to calculate 
the renormalized Uehling-potential generated exclusively by
the induced Wichmann-Kroll charge density 
\begin{eqnarray}
\label{eq:25}
{\cal U}^{{\rm VPVPb}}_{{\cal F}1,{\rm WK}}(r) &=& 
\int_0^\infty\,{\rm d}r'\,4\pi r'^2\,
\left[{\rm e}\rho^{{\rm WK}}_{{\rm ren}}(r')\right]\,f(r,r')\quad ,
\end{eqnarray}
and the renormalized Wichmann-Kroll-type Potential 
${\cal U}^{{\rm VPVPb}}_{{\cal F}2}$ according to the subtraction
scheme (\ref{eq:24}).

As a first step towards the evaluation of the energy correction (\ref{eq:26}) 
we consider the contribution
\begin{eqnarray}
\label{eq:27}
\Delta E_A^{{\rm VPVPb}}({\cal F}1,{\rm WK}) &=& \langle\phi_A|
{\cal U}^{{\rm VPVPb}}_{{\cal F}1,{\rm WK}}|\phi_A\rangle \quad .
\end{eqnarray}
This correction to the Lamb shift of the bound state $\phi_A$ is related to
the change of the Uehling potential Eq. (\ref{eq:25}) arising 
from the Wichmann-Kroll part of the induced vacuum polarization. 
At first we wish to derive an estimate for this 
correction to the Lamb shift of the ground state in hydrogen-like Lead
and Uranium.

The Wichmann-Kroll charge density $\rho^{{\rm WK}}_{{\rm ren}}$ is calculated based on
the partial wave decomposition of the Coulomb propagator and of the free
propagator as developed in Ref. \cite{sam88}. It is obtained from
\begin{eqnarray}
\label{eq:28}
{\rm e} \rho^{{\rm WK}}_{{\rm ren}}(r) &=& \frac{\alpha}{\pi} \int_0^\infty
\frac{{\rm d}u}{2\pi}\,\sum_{|\kappa|=1}^{\infty}\,|\kappa|\,\Re \left\{
\sum_{i=1}^2 \,G^{ii}_\kappa (r,r,{\rm i}u)\right. \nonumber \\
&&\left. \hspace{4cm} 
+ \int_0^\infty {\rm d}r'\,r'^2\,V^{\rm C}(r')\,
\sum_{i,j=1}^2\,\left[F^{i,j}_\kappa (r,r',{\rm i}u)\right]^2\right\}
\quad , 
\end{eqnarray}
where the summation over $\kappa$ is terminated at some maximum value
$|\kappa_{{\rm max}}|$.
The quantities $G^{ii}_\kappa$ and $F^{i,j}_\kappa$ denote the partial 
wave decompositions of the free and the bound propagators. 
Apart from the long-range tail where the 
Wichmann-Kroll vacuum polarization charge density is positive, a 
strongly pronounced maximum of negative charge density occurs in the vincinity 
of the nuclear surface (see Ref. \cite{sam88}). 
Integrating over this $r$-range one obtains a total induced negative charge 
\begin{eqnarray}
\label{eq:29}
Q^{{\rm WK}}_- &=& {\rm e} Z^{{\rm WK}} = \int_0^{r_-} {\rm d}r\,4\pi\,
\rho^{{\rm WK}}_{{\rm ren}}(r) \quad .
\end{eqnarray}
Since the Wichmann-kroll density is strongly localized near the nuclear
surface it acts almost like an additional negatively charged spherical
shell surrounding the nucleus. A strongly bound electron experiences the
reduced nuclear charge. This suggests replacing the corresponding
potential ${\cal U}^{{\rm VPVPb}}_{{\cal F}1,{\rm WK}}$ by the  
Uehling potential generated by  a spherical shell carrying the negative 
charge $Q^{{\rm WK}}_-$ as an approximation. In this case the integral
Eq. (\ref{eq:25}) can be evaluated imediately. If we further employ a 
spherical shell model for the nuclear charge distribution, a simple 
scaling-law is derived.   
This relates the energy shift Eq. (\ref{eq:27}) with the first-order
Uehling correction $\Delta E^{{\rm VP}}_A({\rm Ueh})$ according to:

\begin{eqnarray}
\label{eq:30}
\Delta E_A^{{\rm VPVPb}}({\cal F}1,{\rm WK}) &\simeq& \frac{Z^{{\rm WK}}}{Z}\,
\Delta E_A^{{\rm VP}}({\rm Ueh}) \quad .
\end{eqnarray}
The validity of this scaling-law is supported by the fact that the 
numerical results for the Uehling correction do not depend significantly upon details of
the extended nuclear charge distribution.
With the scaling-law at hand we have an additional tool for testing the
numerical results for the Wichmann-Kroll-in-Uehling contribution 
Eq. (\ref{eq:27}) for strongly bound electrons. 

\begin{figure}
\centerline{\mbox{\epsfxsize=10cm\epsffile{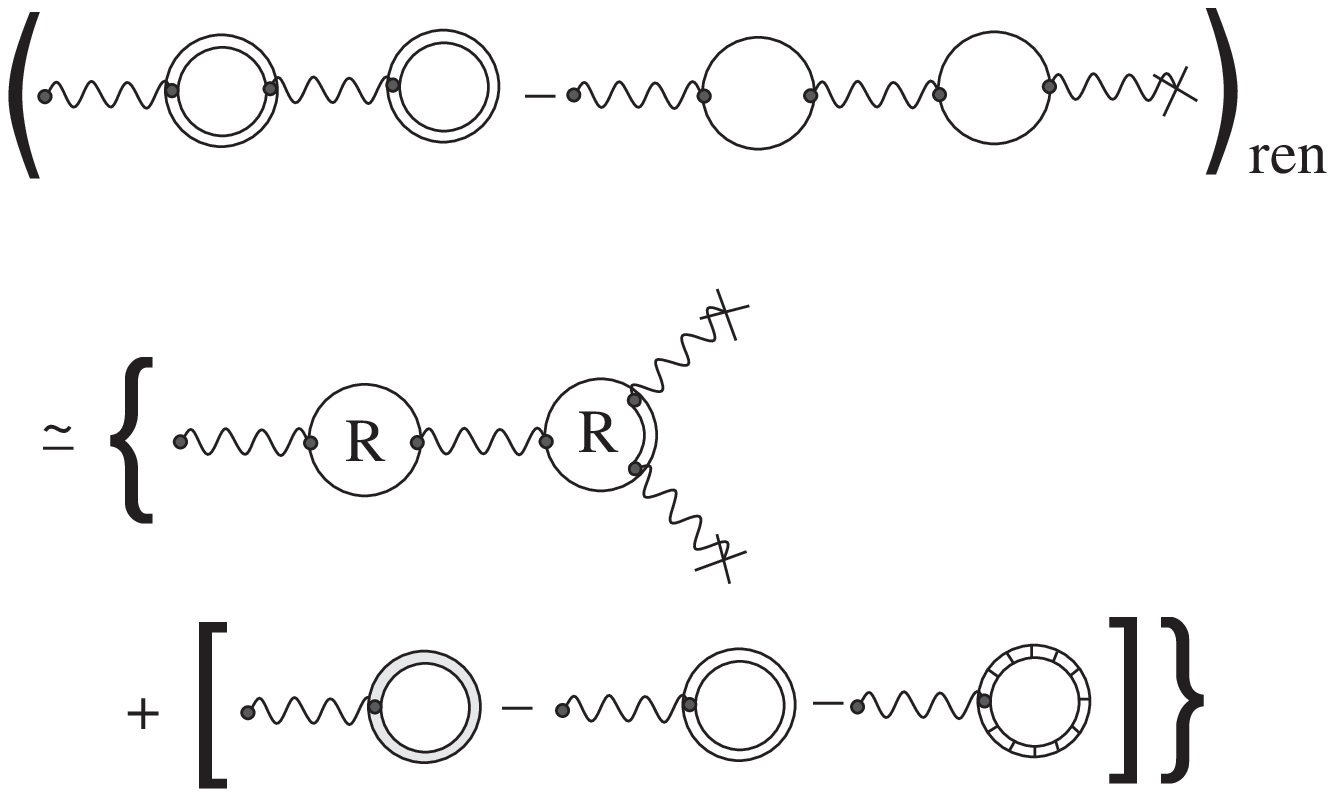}}}
\caption{\label{f:4b} Diagrammatic representation the subtraction scheme
for deducing the higher-order in ($Z\alpha$) part of the 
renormalized two-loop vacuum polarization potential 
${\cal U}^{{\rm VPVPb}}_{{\rm ren}}$. The first term in curly brackets 
stands for the part ${\cal U}^{{\rm VPVPb}}_{{\cal F}1,{\rm WK}}$ and 
the second term in square brackets represents the
Wichmann-Kroll-type potential ${\cal U}^{{\rm VPVPb}}_{{\cal F}2}$
according to Eq. (\protect\ref{eq:24b}). The third one-loop potential
involves Dirac states in the presence of the vacuum polarization potential
$V^{{\rm VP}}_{{\rm ren}}$ only.}
\end{figure}

The complete evaluation of the Wichmann-Kroll-type contribution 
\begin{eqnarray}
\label{eq:31}
\Delta E_A^{{\rm VPVPb}}({\cal F}2) &=& \langle\phi_A|
{\cal U}^{{\rm VPVPb}}_{{\cal F}2}|\phi_A\rangle \quad .
\end{eqnarray}
will be performed numerically according to the subtraction scheme
introduced by Eq. (\ref{eq:24}) respectively (\ref{eq:24b}). Fig. \ref{f:4b}
illustrates the subtraction scheme we applied for numerical calculations.

\section{Results and discussion}
The scaling-law Eq. (\ref{eq:30}) derived above for the ''Wichmann-Kroll in
Uehling'' part $\Delta E_A^{{\rm VPVPb}}({\cal F}1,{\rm WK})$ may  
only be useful as test for its complete numerical calculation.   
In order to employ the scaling-law the first-order Uehling correction
$E_A^{{\rm VP}}({\rm Ueh})$ and the Wichmann-Kroll charge density
$\rho^{{\rm WK}}$ need to be determined. This can be achieved by means
of very accurate numerical procedures developed earlier (see e.g. 
\cite{sam88,pea93,bea97} and references therein). For a discussion of  
technical details encountered with the evaluation of 
$\rho^{{\rm WK}}_{{\rm ren}}$ we
refer to \cite{bea97}. The total induced negative charge number 
$Z^{{\rm WK}}_{{\rm ren}}$ 
is obtained by integrating the Wichmann-Kroll density from
the origin $r=0$ up to $r=r_-$ where $\rho^{{\rm WK}}_{{\rm ren}}$ 
changes its sign.
Table \ref{t:1} gives numerical results for the first-order Uehling 
correction, the induced negative charge number and for the corresponding 
energy shifts $\Delta E_A^{{\rm VPVPb}}({\cal F}1,{\rm WK})$ according to 
the scaling-law for the $1$S-ground state of hydrogen-like Lead and Uranium
respectively. The energy correction is repulsive, since the induced charge 
is negative. 

Let us compare the effect of the higher-order contribution
to the two-loop correction $\Delta E^{{\rm VPVPb}}_A({\rm h.o.})$ 
defined by Eq. (\ref{eq:26}) with the one obtained in Uehling approximation
$\Delta E^{{\rm VPVPb}}_A({\cal F}1,{\rm Ueh})$. The latter is tabulated in
the first column of 
Table \ref{t:2} for K- and L- shell electrons in hydrogen-like Lead and
Uranium. This Uehling-in-Uehling contribution is attractive and
amounts to less than $20$\% of the total K\"all\'{e}n-Sabry correction 
to the Lamb shift as tabulated in \cite{sgs93}. 
The results for the exact two-loop correction are tabulated in the second
column of Table \ref{t:2}.

\begin{figure}
\centerline{\mbox{\epsfxsize=10cm\epsffile{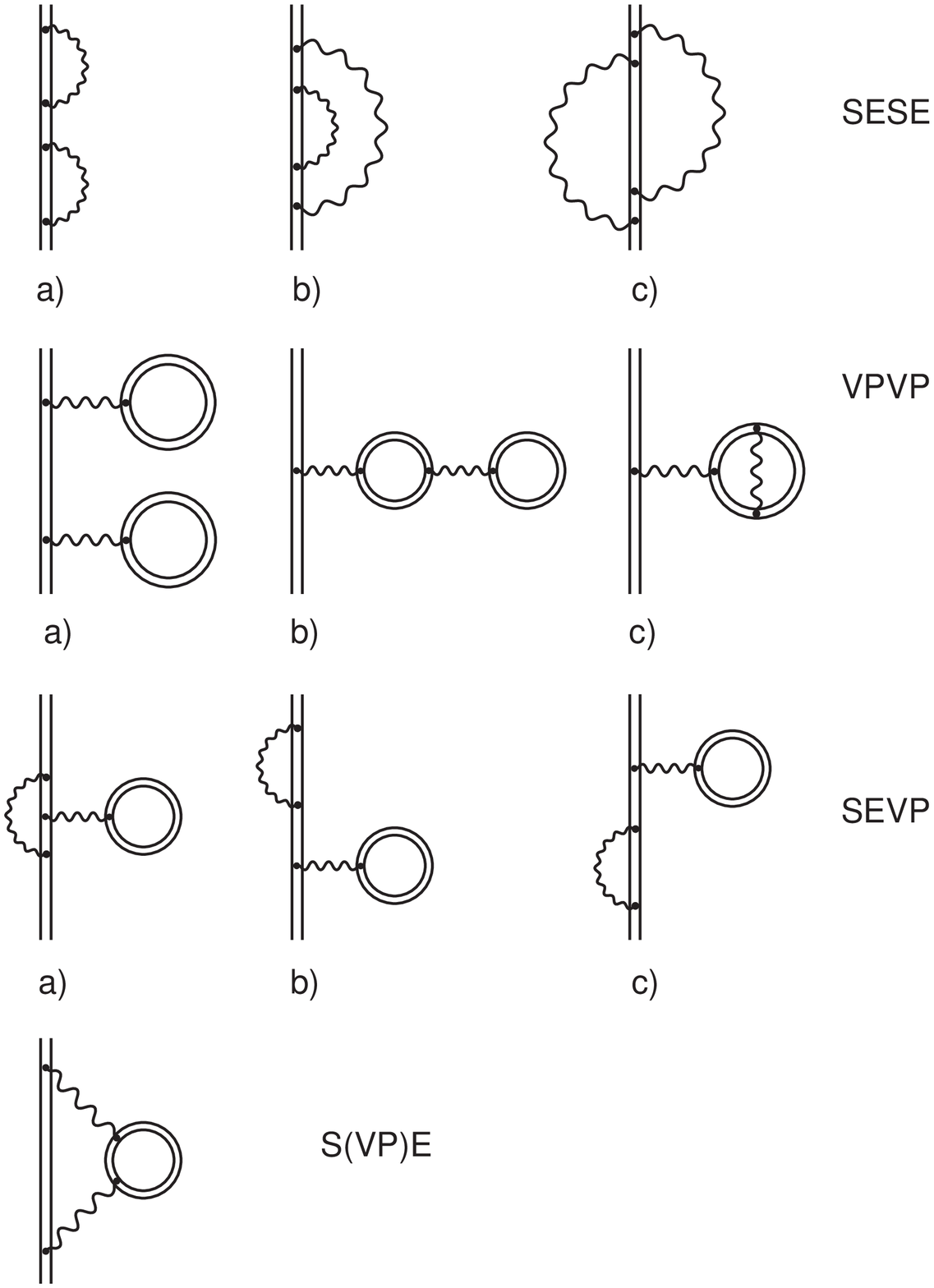}}} 
\caption{\label{f:1} QED corrections of order $\alpha^2$ in hydrogen-like
atoms. The double lines indicate wave functions and propagators in the
external Coulomb field of the nucleus.}
\end{figure}

In comparison with the exact numerical results for the 
Wichmann-Kroll-in-Uehling contribution as presented in the third column 
of Table \ref{t:2} the scaling-law indeed gives the right order of magnitude
but leads to a systematic overestimation due to the neglection of the
long-range tail of $\rho^{{\rm WK}}_{{\rm ren}}$. 
The numerical results for the Wichmann-Kroll-type contribution
$\Delta E_A^{{\rm VPVPb}}({\cal F}2)$, which have been obtained according 
to the representation (\ref{eq:24b}),
are presented in the last column of Table \ref{t:2}. 
As for the Uehling-in-Uehling correction this contribution carries the
same overall sign, i.e. it also acts attractive. 
It comprises about $40$ \%
of the two-loop diagram evaluated in Uehling approximation. This result
also indicates that higher-order ($Z\alpha$) contributions to the one-loop
polarization insertion of the photon propagator may be not small compared
with the Uehling approximation. Accordingly, it may be not surprising at all,
if the higher-order contribution to the S(VP)E - correction (see 
Fig. \ref{f:1}) may turn out to be important as well. 

Even in the case of Uranium the order of magnitude for the total 
higher-order part $\Delta E_A^{{\rm VPVPb}}({\rm h.o.})$ (see Table
\ref{t:2}) is about $\sim 10^{-2}$ eV for the 1S-state,
which indicates that the effect of the higher-order contributions to the 
two-loop vacuum polarization alone are far too small to be detected via
Lamb-shift measurements with current accuracies. Furthermore, this correction
is about one order of magnitude below the natural limitations for tests
of QED set by nuclear polarization effects \cite{np}. It might also be 
instructive to compare this effect with the uncertainties of the 
$1$S-energy level 
caused by the uncertainties in the determination of the nuclear radii.
The rms-radius for Uranium is given by $<r^2>^{1/2}_{{\rm U}} = 5.8604(23)$ 
fm \cite {zea84} and leads to an uncertainty of
about $\delta E^{{\rm U}}_{1s} \sim 0.1$ eV. 
Although a complete numerical evaluation of the exact K\"all\'{e}n-Sabry
diagrams, i.e., the higher-order contributions of the
self-energy-corrected one-loop diagram VPVPc remain to be performed, 
the goals of this paper have been achieved: 1. We have presented a   
calculational scheme for evaluating the two-loop vacuum polarization
corrections to all orders in the interaction with the external Coulomb
potential. The problem is reduced to the evaluation of an effective 
one-loop correction. Thus, the partial wave decomposition can be employed in a 
similar way as it has been used successfully in calculations of the ordinary 
first-order vacuum polarization correction. 2. Numerical results for exact
two-loop vacuum polarization correction
to the Lamb shift of K- and L-shell electrons in hydrogen-like 
Lead and Uranium have been obtained. 
It turns out to be below the natural limits set by nuclear 
polarization effects and by the uncertainties of nuclear parameters.
3. Although the effect of the higher-order contributions to the two-loop
diagram turn out to be small a further uncertainty in Lamb-shift calculations
has been eliminated.
Aiming for a relative precision of $10^{-6}$ of theoretical predictions
of the binding energy we conjecture that the still unknown exact 
two-photon self energy corrections remain the major source of uncertainties.

\acknowledgments

The authors are grateful to Sten Salomonson and Per Sunnergren 
for interesting discussions. Financial support has been provided 
by the BMBF, the DFG, the DAAD, the SI, and the GSI (Darmstadt).

\newpage
\clearpage
\begin{table}
\caption{\label{t:1} Estimate of the ''Wichmann-Kroll in Uehling'' 
contribution to the $1$S$_{1/2}$-Lamb shift in hydrogen-like Lead and
Uranium. The values given for the first-order Uehling correction
$E_A^{{\rm VP}}({\rm Ueh})$ are calculated assuming a uniform sphere model
for the nuclear charge density \protect\cite{pea96}. The Wichmann-Kroll 
density $\rho^{{\rm WK}}_{{\rm ren}}$ is calculated for a spherical shell model
\protect\cite{bea97}.}

\bigskip
\begin{tabular}{cddd}
system & $\Delta E_{1{\rm S}_{1/2}}^{{\rm VP}}({\rm Ueh})$ [eV] & $Z^{{\rm WK}}$ &
$\Delta E_{1{\rm S}_{1/2}}^{{\rm VPVPb}}({\cal F}1,{\rm WK})$ [eV] \\ 
\hline \hline
$^{238}_{\phantom{x}92}$U  & --93.58  & --0.006 & 0.0061 \\
$^{208}_{\phantom{x}82}$Pb & --50.70  & --0.004 & 0.0024  
\end{tabular}
\end{table}

\clearpage
\begin{table}
\caption{\label{t:2} The results for the two-loop vacuum polarization
contribution in Uehling approximation 
$\Delta E_A^{{\rm VPVPb}}({\cal F}1,{\rm Ueh})$ (Eq. (\protect\ref{eq:26b}))
in comparison with the exact two-loop correction $\Delta E_A^{{\rm VPVPb}}$
to the Lamb shift of strongly bound electrons in hydrogen-like Lead and
Uranium. The higher-order contributions 
$\Delta E_A^{{\rm VPVPb}}({\cal F}1,{\rm WK})$ (Eq. (\protect\ref{eq:30})) 
and $\Delta E_A^{{\rm VPVPb}}({\cal F}2)$ (Eq. (\protect\ref{eq:31})) are 
listed separately. Energies are given in units of eV.}

\bigskip
\begin{tabular}{ccdddd}
system & state & $\Delta E_A^{{\rm VPVPb}}({\cal F}1,{\rm Ueh})$ &
$\Delta E_A^{{\rm VPVPb}}$ & $\Delta E_A^{{\rm VPVPb}}({\cal F}1,{\rm WK})$ &
$\Delta E_A^{{\rm VPVPb}}({\cal F}2)$ \\  \hline
 & $1s_{1/2}$ & --0.1150 & --0.1530 & 0.0040    & --0.0420   \\
$^{238}_{\phantom{x}92}$U
 & $2s_{1/2}$ & --0.0220 & --0.0286 & 0.00074   & --0.0073 \\
 & $2p_{1/2}$ & --0.0023 & --0.0036 & 0.000079  & --0.0014 \\ 
                           \hline
 & $1s_{1/2}$ & --0.0520 & --0.0685 & 0.0015   & --0.0180   \\ 
$^{208}_{\phantom{x}82}$Pb  
 & $2s_{1/2}$ & --0.0092 & --0.0118 & 0.00027  & --0.0029  \\ 
 & $2p_{1/2}$ & --0.0006 & --0.0010 & 0.000018 & --0.0004  \\ 
\end{tabular}
\end{table}

\newpage
\clearpage

\clearpage

\clearpage

\clearpage

\clearpage

\clearpage


\begin{thebibliography}{99}
\bibitem{sea91} J. Schweppe, A. Belkacem, 
             L. Blumenfeld, N. Claytor, B. Feinberg,
             H. Gould, V.~E. Costram, 
             L. Levy, S. Misawa, J.~R. Mowat, and M.~H. Prior,
             Phys. Rev. Lett. {\bf 66}, 1434 (1991).
\bibitem{bey94}
             H. F. Beyer, D. Liesen, 
             F. Bosch, K. D. Finlayson, M. Jung, O. Kleppner,
             R. Moshammer, K. Beckert, H. Eickhoff, B. Franzke, F. Nolden,
             P. Sp\"adtker, M. Steck, G. Menzel, R. D. Deslattes,
             Phys. Lett. {\bf A184}, 435 (1994).
\bibitem{sea93}
             Th.~St\"{o}hlker, P.~H.~Mokler, K.~Beckert, F.~Bosch,
             H.~Eickhoff, B.~Franzke, M.~Jung, T.~Kandler, 
             O.~Klepper, C.~Kozhuharov, R.~Moshammer, F.~Nolden,
             H.~Reich, P.~Rymuza, P.~Sp\"adtke, and M.~Steck,
             Phys.~Rev.~Lett. {\bf 71}, 2184 (1993).
\bibitem{bey95} 
             H.~F.~Beyer, IEEE Trans.~Instrum.~Meas. {\bf 44}, 510 (1995);\\
             H.~F.~Beyer, G.~Menzel, D.~Liesen, A.~Gallus,
             F.~Bosch, R.~Deslattes, P.~Indelicato, Th.~St\"ohlker,
             O.~Klepper, R.~Moshammer, F.~Nolden, H.~Eickhoff,
             B.~Franzke, and M.~Steck, Z.~Phys.~{\bf D 35}, 169 (1995).
\bibitem{eides}
            M.~I.~Eides and H.~Grotch, D.~A.~Owen, 
            Phys.~Lett. {\bf B294}, 115 (1992);\\
            M.~I.~Eides and H.~Grotch, Phys.~Lett. {\bf B301}, 127 (1993);\\
            M.~I.~Eides and H.~Grotch, Phys.~Lett. {\bf B308}, 389 (1993);\\
            M.~I.~Eides, S.~G.~Karshenboim, and V.~A.~Shelyuto,
            Phys.~Lett.~{\bf B 312}, 389 (1993);\\
            M.~I.~Eides, H.~Grotch, and P.~Pebler, 
            Phys.~Rev. {\bf A50}, 144 (1994).\\
\bibitem{pachucki}
            K.~Pachucki, Phys.~Rev. {\bf A48}, 2609 (1993);\\
            K.~Pachucki, Phys.~Rev.~Lett. {\bf 72}, 3154 (1994).
\bibitem{pea96}
            H.~Persson, I.~Lindgren, L.~Labzowsky, T.~Beier,
            G.~Plunien, and G.~Soff, Phys.~Rev.~{\bf A 54}, 2805 (1996).
\bibitem{bea97}
            T.~Beier, M.~Greiner, G.~Plunien, and G.~Soff, 
            accepted for publication in J. Phys. {\bf B}, (1997).
\bibitem{bas88}
            T.~Beier and G.~Soff, Z.~Phys.~{\bf D8}, 129 (1988). 
\bibitem{sgs93}
            S.~M.~Schneider, W.~Greiner, and G.~Soff, 
            J.~Phys. {\bf B26}, L529 (1993).
\bibitem{kas55} 
            G.~K\"all\'en and A.~Sabry,
            Mat.~Fys.~Medd.~Dan.~Vid.~Selsk. {\bf 29}, 17 (1955)
\bibitem{lam95} 
            L.~Labzowsky and A.~O.~Mitrushenkov, 
            Phys. Lett. {\bf A198}, 333 (1995); \\
            L.~Labzowsky and A.~O.~Mitrushenkov, 
            Phys. Rev. {\bf A53}, 3029 (1996). 
\bibitem{sam88}
            G.~Soff and P.~J.~Mohr, Phys.~Rev.~{\bf A 38} (1988) 5066.
\bibitem{pwr} 
            H.~Persson, I.~Lindgren, and S.~Salomonson,
            Phys. Scr. {\bf T46}, 125 (1993); \\
            I.~Lindgren, H.~Persson, S.~Salomonson, and A.~Ynnermann,
            Phys. Rev. {\bf A47}, 4555 (1993). 
\bibitem{pea93}
            H.~Persson, I.~Lindgren, S.~Salomonson, and P.~Sunnergren,
            Phys. Rev. {\bf A48}, 2772 (1993).

\bibitem{np} 
G. Plunien, B. M\"uller, and W. Greiner, Phys. Rev. {\bf A43}, 5853 (1991);\\
A. V. Nefiodov, L. N. Labzowsky, G. Plunien, and G. Soff, 
Phys. Lett. {\bf A188}, 371 (1996);\\
A. V. Nefiodov, L. N. Labzowsky, G. Plunien, T. Beier, and G. Soff, 
J. Phys. {\bf B29}, 3841 (1996).
            
\bibitem{zea84}
V. Zacek, H. Bohn, H. Brum, F. von Feilitzsch, G. Giorginis, P. Kienle,
S. Schuhbeck, Z. Phys. {\bf A318}, 7 (1984).

\end{thebibliography}
\end{document}